\documentclass[twocolumn,showpacs,amsmath,amssymb]{revtex4}
\usepackage{graphicx}
\usepackage{dcolumn}
\usepackage{bm}
\newcommand{\be}{\begin{eqnarray}}
\newcommand{\en}{\end{eqnarray}}
\newcommand{\ben}{\begin{eqnarray*}}
\newcommand{\enn}{\end{eqnarray*}}

\newcommand{\bi}{\begin{itemize}}
\newcommand{\ei}{\end{itemize}}

\def\prl#1#2#3{{ Phys.   Rev.   Lett.  } {\bf #1}, #2 (#3)}

\def\pre#1#2#3{Phys.   Rev.   E {\bf #1}, #2 (#3)}

\def\jsp#1#2#3{J.   Stat.   Phys.   {\bf #1}, #2 (#3)}

\def\rmp#1#2#3{Rev.   Mod.   Phys.   {\bf #1}, #2 (#3)}
\def\epl#1#2#3{Europhys. Lett. {\bf #1}, #2 (#3)}

\def\jsm#1#2#3#{J.Stat.Mech. {\bf#1}, #2 (#3)}
\begin{document}
\title{Is Turbulence as Simple as Tossing a Coin?}
\author{Jayanta Kumar Bhattacharjee}
\email{jkb@bose.res.in}
\affiliation{S.N. Bose National Centre for Basic Sciences, Saltlake, Kolkata 700098, India}
\author{Sagar Chakraborty}
\email{sagar@bose.res.in}
\affiliation{S.N. Bose National Centre for Basic Sciences, Saltlake, Kolkata 700098, India}
\author{Arnab Saha}
\email{arnab@bose.res.in}
\affiliation{S.N. Bose National Centre for Basic Sciences, Saltlake, Kolkata 700098, India}
\date{\today}
\begin{abstract}
A large variety of problems in statistical physics use a Gaussian distribution as a starting point. For the problem of intermittency in fluid turbulence, the Gaussian approximation is not a useful beginning. We find that the Cramer's rate function in the theory of large deviations as used in a simple coin toss is a promising starting point for giving an account of intermittency. In addition, it offers another view of Jarzynski equality.
\end{abstract}
\maketitle
%
\indent Large deviations play a significant role in many branches of non-equilibrium statistical physics\cite{1,2}. They are difficult to handle because their effects though small, are not amenable to perturbation theory. All the conventional perturbation theories in statistical physics are fashioned about a Gaussian distribution, which almost by definition, is the distribution with no large deviations. This can be seen in static critical phenomena, critical dynamics, dynamics of interfacial growth, statistics of polymer chain and myriad other problems\cite{3}. However, the Gaussian model fails to be a starting point while discussing intermittency in fluid turbulence\cite{Frisch,She,4,5,6,7,8,9,10,11,12,13,15}. In the large deviation theory, the central role is played by the distribution associated with tossing of a coin. Our contention is: the simple coin toss is the ``Gaussian model" of problems where rare events play significant role. We illustrate this by applying it to the studies of intermittency in fully developed turbulence and Jarzynski equality.
\\
\indent The high Reynolds number turbulence remains the prime age old problem dominated by rare events, which still eludes a satisfactory theoretical understanding. Before we plunge into the problem of modeling intermittency in a turbulent fluid, let us begin by briefly reviewing the large deviation theory in the context of a coin toss experiment\cite{17}.
Suppose we have a biased coin, such that for each toss the probability of obtaining ``head" is `$p$'. If we assign the value $1$ to the outcome ``head" (each outcome is denoted by $X_i$ where $i=1,2,...$) and $0$ to the outcome ``tail", then the mean after $N$ trials is 
\begin{equation}
M_N=\frac{1}{N}\sum_{i=1}^N X_i
\label{1}
\end{equation}

\noindent As $N\rightarrow\infty$, it is expected that $M_N\rightarrow p$. The question is: For large N, what is the probability that $M_N$ differs from $p$ by at least $x$ (where $x$ is any pre-assigned fraction less than unity)? The meaning of large deviation is that however large $N$ may be this probability is nonzero and if the $X_i$'s are bounded, independent and identically distributed random variables, then Crammers theorem asserts that the tail of the probability distribution of $X_i$ is given by
\begin{eqnarray}
\left.\begin{array}{cc}
P(M_N>x)\approx e^{-NI(x)}& \phantom{xx}\textrm{for}\phantom{x} x>p \\
P(M_N<x)\approx e^{-NI(x)}& \phantom{xx}\textrm{for}\phantom{x} x<p
\label{2}
\end{array}\right\}
\end{eqnarray}

\noindent To apply this result in different disciplines of statistical physics, we require $P(M_N \approx x)$ and it is Varadhan's theorem that ensures that the sequence $M_N$ itself satisfies a large deviation principle i.e. $P(M_N \approx x)\sim e^{-NI(x)}$. For the coin toss under consideration, Chernoff's formula gives the rate function $I(x)$ as follows:
\begin{equation}
I(x)=x\ln\frac{x}{p}+(1-x)\ln\frac{1-x}{1-p}
\label{3}
\end{equation}
\noindent and this is the central result that we will use.
\\
\indent Turning to turbulence, in 1941 Kolmogorov\cite{Kolmogorov} invoked the concept of Richardson's cascade\cite{19} of eddies to propose a phenomenological model (K41) for three dimensional incompressible turbulence at high Reynolds number. Even today this is the cornerstone of our understanding of turbulence. Understanding turbulence is understanding the small scale behaviour of the velocity structure function $S_q(l)$, where $S_q(l)\equiv\langle|\Delta\vec{v}.(\vec{l}/|l|)|^q\rangle$, with $\Delta \vec{v}\equiv\vec{v}(\vec{r}+\vec{l})-\vec{v}(\vec{r})$ and `$l$' is a distance which is short compared to macroscopic length scales like the system size but is large compared to molecular scale where viscous dissipation takes place. The angular bracket denotes ensemble average (i.e. average over different values of `$\vec{r}$'). The observation is that $S_q(l)$ has a scaling behaviour $l^{\zeta_q}$ where $l$ is in the range indicated (so called inertial range). Finding  $\zeta_q$ can be described as the holy grail of turbulence. K41 gives $\zeta_q=q/3$ --- a result which is exact for $q=3$ and very close to experimental findings for low value of $q$. There is systematic departure from $q/3$ at relatively higher values of $q$. This is the phenomenon of intermittency. Of particular interest is the case $q=6$. Since $|\Delta v|^3/l$ is a measure of the local energy transfer rate (same as energy input and energy dissipation rate in K41 and thus a constant), we expect $\zeta_6=2$. The deviation $2-\zeta_6$ is thus a very sensitive quantity and is often singled out for special treatment. The exponent $\mu=2-\zeta_6$ is formally called the intermittency exponent and the experimental measurements agree on a value 0.2 for $\mu$. It can be viewed as the co-dimension of dissipative structures.
\\
\indent The model of intermittency are usually constructed on a phenomenological basis by thinking of various ways of modifying the Richardson's cascade picture. The $\beta$-model, the bifractal model and the multifractal model all belong to this class. The crucial hypothesis is that the daughter eddies produced from the mother eddies are not space filling and the active part of space is in general a multifractal. The velocity field has different scaling exponents on different fractal sets that form the multifractal structure. These scaling exponents can, in principle, yield $\zeta_q$. This multifractality can also be defined and measured in terms of the fluctuations of the local dissipation rate rather than in terms of the fluctuations of the velocity increments $\Delta{v}$. The key element, that is needed to define multifractality in terms of dissipation is the local space average of energy dissipation over a ball of radius $l$ centered around a point at $\vec{r}$: $\varepsilon_l(\vec{r})\equiv\frac{3\nu}{8\pi l^3}\int_{|\vec{r^{\prime}}-\vec{r}|<l}d^3\vec{r^{\prime}}\sum_{i,j}[\partial_jv_i(\vec{r^{\prime}})+\partial_iv_j(\vec{r^{\prime}})]^2$. If the dissipation is multifractal, moments of $\varepsilon_l$ follow a power law behaviour at small $l$, i.e. $\langle\varepsilon_l^q\rangle\sim l^{\tau_q}$. Kolmogorov's refined similarity hypothesis relates the statistical properties of fluctuation of velocity increment to those of the space averaged dissipation and yields: $\zeta_q=\frac{q}{3}+\tau_{q/3}$. We now carry out the usual speculation that since the higher order velocity structure factors differ most strongly from K41, then the probability distribution for the velocity increments must differ most strongly from that appropriate to K41 in the tail of the distribution. The tail of a distribution involves rare events and this is how the theory of large deviations enters the picture. Following Landau's observation on K41\cite{Landau}, Kolmogorov\cite{4} and Obukhov\cite{5} introduced fluctuations in the dissipation rate. Careful experiments revealed the existence of these fluctuations. The fluctuations, however, occur rarely and these are the rare events of turbulence. This allows us to establish a quantitative bridge between turbulence and theory of large deviations.
\\
\indent More than a decade ago, Stolovitzky and Sreenivasan\cite{15}, in somewhat different approach tried to validate refined similarity hypothesis by viewing turbulence as a general stochastic process (fractional Brownian motion to be precise). While this was a very significant achievement, there was a shortcoming in that the theory ruled out the existence of correlation functions like $S_3$. It indeed is surprising since the readers may know that the only exact non-trivial result existing in the theory of turbulence is Kolmogorov law: $S_3(l)=-\frac{4}{5}\varepsilon l$. However as we shall note, their approach allows us to make direct contact with the terms of large deviation that signify the occurrence of rare events. It can be observed that $\varepsilon_l$ plays the role of $M_N$ of equation (\ref{1}) and it is the deviation from the expected mean $\varepsilon$ that we are interested in. As $l\rightarrow \infty$, this deviation variable has a distribution according to the role of equation (\ref{2}). We hope a simplification: The $\varepsilon_l-\varepsilon$ can range from large negative to large positive values. We bring the range between $0$ to $1$ by defining a variable as:
\begin{equation}
Z_T(\varepsilon_l)\equiv\frac{1}{2}\left[1+\tanh\left(\frac{\varepsilon_l-\varepsilon}{\Xi}\right)\right]
\label{Z}
\end{equation}
\noindent where $\Xi$ is a constant with dimension of $\varepsilon$. We now make the drastic assumption that since $\varepsilon_l-\varepsilon$ is a rare event, the distribution of $Z_T$ can be considered similar to that for the coin-toss with a biased coin and accordingly, we can hypothesize that
\begin{equation}
P(Z_T)\propto e^{-NI(Z_T)}
\label{PZ}
\end{equation}
\noindent Here, $N$ is number of random variables. This simple model yields value of $\mu\approx 0.16$ which is quite close to the presently accepted value. Also, a $\zeta_q$ vs $q$ plot has been obtained that is not only convex but also follows She-Leveque scaling\cite{She} faithfully enough for a model as simple as this. Please refer to figure-1.
In what follows we describe how these results are arrived at.
\\
\indent The one dimensional velocity derivative can be use to express the global average of the full energy dissipation if local isotropy exists\cite{29,30}. The velocity increment is given by
\begin{equation}
\Delta v(l)=\int_r^{r+l}\frac{dv}{dr}dr
\label{m1}
\end{equation}
\noindent and {\it ergo,} the energy dissipation rate is
\begin{equation}
\varepsilon(l)=\frac{15\nu}{l}\int_r^{r+l}\left(\frac{dv}{dr}\right)^2dr
\label{m2}
\end{equation}
\noindent If we define $D_i\equiv\left.\frac{dv}{dr}\right|_i\left[\frac{\eta\sqrt{15\varepsilon}}{\left(\eta\varepsilon\right)^{1/3}}\right]$ and $N\equiv\frac{l}{K\eta}$ (where, $\eta$ is Kolmogorov scale, ${\left(\eta\varepsilon\right)^{1/3}}$ is Kolmogorov velocity scale and $K$ is the number of Kolmogorov scales over which one obtains smoothness), then equation (\ref{m2}) may be rewritten, upon discretization, as:
\begin{equation}
\varepsilon_l-\varepsilon=\frac{1}{N}\sum_{i=1}^{N}Y_i
\label{m3}
\end{equation}
\noindent Here, $Y_i\equiv D_i^2-\varepsilon$. In the letter, we have assumed the relation (\ref{m3}) to be the parallel of equation (\ref{1}). Owing to the contraction principle, the rate function for $\varepsilon_l-\varepsilon$ and $Z(\varepsilon_l)$ are same. Thus, using equations (\ref{3}), (\ref{Z}) and (\ref{PZ}), we can write:
\begin{equation}
\left\langle\left|\varepsilon_l-\varepsilon\right|^q\right\rangle=\left|\frac{\Xi}{2}\right|^q\left[\frac{\int_0^1\left|\ln\left(\frac{x}{1-x}\right)\right|^q\left\{\left(\frac{p}{x}\right)^x\left(\frac{1-p}{1-x}\right)^{1-x}\right\}^Ndx}{\int_0^1\left\{\left(\frac{p}{x}\right)^x\left(\frac{1-p}{1-x}\right)^{1-x}\right\}^Ndx}\right]
\label{m4}
\end{equation}
\noindent We assume that to the leading order $\left\langle\left|\varepsilon_l-\varepsilon\right|^q\right\rangle\sim l^{\tau_q}$. By trial and error, we fix the inertial range as $N=30$ to $60$ and calculate numerically $\mu(=-\tau_2)=0.16$. Similarly, we calculate $\zeta_q(=q/3+\tau_{q/3})$ for various $q$. Note that to obtain the numerical solution for the integrals in equation (\ref{m4}), we have dropped the diverging terms from the finite series that represent the integrands as they are suitably discretized for their evaluation by Simpson's one-third rule.
\begin{figure}
\includegraphics[width=0.45\textwidth]{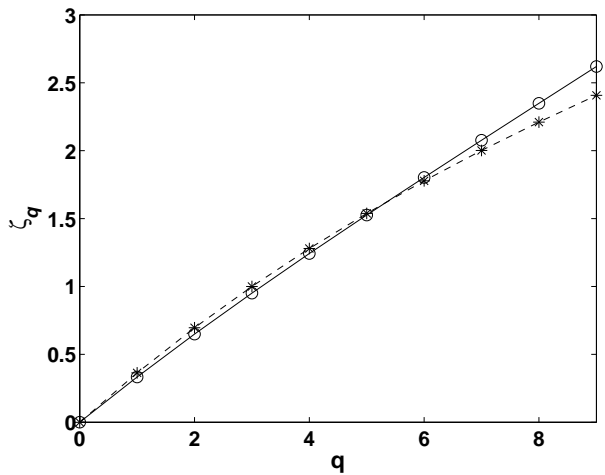}
\caption{{\bf $\zeta_q$ {\it vs.} $q$ curve in fully developed fluid turbulence}. The dashed line joining the asterisk is the celebrated She-Leveque scaling law.The circles joined by the solid line denote the values of $\zeta_p$ (for corresponding $q$) as obtained by dint of the model proposed herein. For every $q$, first $\left\langle\left|\varepsilon_l-\varepsilon\right|^q\right\rangle$ {\it vs.} $N$ is plotted in log-log scale using the data yielded during the numerical integration of equation (\ref{m4}) and then the observation that for $N=30$ to $60$, we get a fairly straight line leads us to attempt fitting the range linearly. The process gives a value for $\tau_q$. The relation $\zeta_q=q/3+\tau_q$, then, tells us what is the corresponding value for $\zeta_q$. One can see, the fit is remarkable. There is room for improvement in extending the inertial range and in getting better fit for higher $\zeta_q$'s. As mentioned in this letter, the form of $Z_T$ is crucial.}
\end{figure}
\\
\indent Our model's inherent bias for the value $0.26$ for the parameter $p$ in order to closely mimic the realistic turbulent fluid's scaling properties would seem so natural when it is compared with a particular successful multifractal cascade model\cite{21} based on a generalized two-scale Cantor set. In that model, as the eddies breakdown into two new ones, the flux of kinetic energy into the smaller scales is hypothesized to be dividing into non-equal fractions $p=0.3$ (quite close to our value of $p=0.26$!) and $1-p=0.7$. It could fit remarkably well the entire spectrum of generalized dimensions\cite{22} and (equivalently) the singularity spectrum (the so-called $f-\alpha$ curve\cite{23}) for the energy dissipation field in many a turbulent flow.
\\
\indent We finally note that the large deviation result that we have used lends insight into yet another frontier of non-equilibrium statistical mechanics --- that of the various fluctuation theorems\cite{24,25,26,27,28}. We will focus on one of them: Jarzynski equality. Here one begins with a system in equilibrium and then switches on a time dependent force  for a period of $\tau$. We consider the different microstates corresponding to an initial equilibrium state and consider the time evolution of each initial condition. The work done $w$ over the time $\tau$ is calculated along each path and then average $e^{-w}$ over the ensemble of all possible initial conditions. This average is denoted by $\left\langle e^{-w}\right\rangle$. This gets related to the exponential of the free energy difference $\Delta F$ between initial and final states leading to Jarzynski's equality. Defining $w_{D}\equiv w-\Delta F$, the equality can also be cast in the form:
\begin{equation}
\left\langle e^{-w_{D}}\right\rangle=1
\label{nnn1}
\end{equation} 
where $w_{D}$ is the dissipative work along a `given' path and the fact that average is unity implies that there are paths for which $w_{D}<0$ --- a case of transient violation of the second law of thermodynamics. These violations can be portrayed as the rare events, which are highlighted by Jarzynski equality.
%
\\
\indent To verify Jarzynski equality {\it i.e.}, in equation (\ref{nnn1}), we consider a system under the application of an explicitly time dependent force that causes the system to evolve from an initial equilibrium state to a final state.
If we repeat the same process $M$ times keeping initial and final microstate fixed, then owing to the inherent stochasticity of the system, every time we get a different dissipative work $w^{j}_{D}$ along a single phase space trajectory ($j$th trajectory) of the `ensemble of $M$ trajectories'.
Let's define a sample mean over the $M$ trajectories as $W_{D}\equiv\frac{1}{M}\sum_{j=1}^M w^{j}_{D}$.
Now, $W_{D}$ can always be expressed as $w_{D}+\frac{1}{M}\sum_{j=1}^M \delta^j$ where $\delta^j$ is the fluctuation around the dissipative work along $jth$ trajectory. As $M$ is large, the summation over all fluctuations will be negligibly small in comparison to the other term.
Therefore, by showing $\left<e^{-W_{D}}\right>=1$, we can basically conclude that $\left<e^{-w_{D}}\right>=1$; hence, Jarzynski equality stands verified. Please refer figure-2.
So, how exactly we prove it?
\\
\indent Well, coming on the technique employed to verify Jarzynski equality, first of all we observe that $W_{D}$ is a random variable measured over the ensemble of all microscopic states.
We define a variable $Z_J(W_D)$ as,
\begin{equation}
Z_J(W_D)=\frac{1}{2}[1-\tanh(W_{D}+c)]
\label{nnn2}
\end{equation}
with an arbitrary $c>0$.
We define: $p\equiv\frac{1}{2}(1-\tanh c)$ and assume that $Z_J(W_D)$ corresponds to a coin toss experiment with a biased coin, where the probability of getting tails is $p$.
We notice that the probability of $W_{D}$ being negative is the same as $Z_J$ having value between $p$ and 1 while $W_{D}$ being positive covers the range $0$ to $p$.
The probability of finding a value of $Z_J$ between $0$ and $1$ is then given by equation (\ref{2}) with the rate function given by the equation (\ref{3}). The biased coin ensures that the negative values of $W_{D}$ --- which basically are the rare events --- are suppressed, as it should be. We note from the definition of $Z_J$ that $e^{-W_{D}}=e^c\sqrt{\frac{Z_J}{1-Z_J}}$ and hence its ensemble average is,
\begin{equation}
\left<e^{-W_{D}}\right>=e^c\frac{\int_0^1\sqrt{\frac{x}{1-x}}\left(\frac{p}{x}\right)^{Nx}\left(\frac{1-p}{1-x}\right)^{N(1-x)}dx}{\int_0^1\left(\frac{p}{x}\right)^{Nx}\left(\frac{1-p}{1-x}\right)^{N(1-x)}dx}
\label{pic2}
\end{equation}
For different values of $p<1/2$, we numerically evaluate the above integral for $N$ number of realizations to reach the conclusions discussed earlier in the letter.
\begin{figure}
\includegraphics[width=0.45\textwidth]{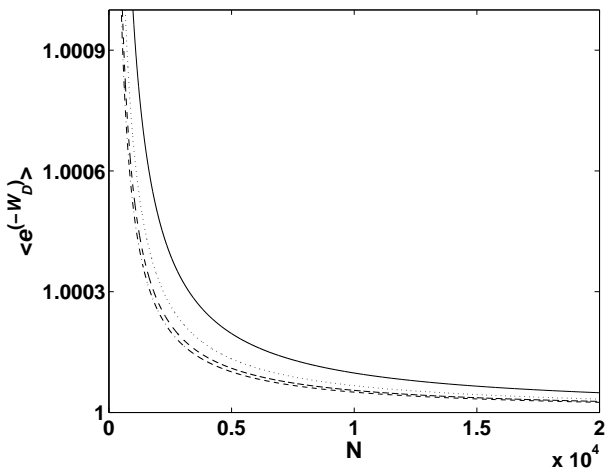}
\caption{{\bf Verification of Jarzynski equality}. The solid, dotted, dashed and dot-dashed lines are respectively for $p=0.15,0.25,0.35$ and $0.45$. The curves in the figure are obtained by numerically integrating equation (\ref{pic2}).}
\end{figure}
\\
\indent In the closing, one would appreciate the simplicity of biased coin-toss experiments and its reasonably astonishing success in predicting $\mu$ renders the need for more complicated models redundant. We believe just by being able to find a more appropriate function $Z_T$, we can make big leaps in the rather complex theory of turbulence.
Moreover, by showcasing that Jarzynski equality can be verified within the similar framework, we inject confidence into our contention that the simple coin toss is the ``Gaussian model" of problems where rare events play significant role.
\\
\\
C.S.I.R. (India) is gratefully acknowledged for financially supporting Sagar Chakraborty by awarding him senior research fellowship.

\end{document}